\documentclass[showpacs,prd,twocolumn,aps]{revtex4}

 \usepackage{bm}
 \usepackage{amsmath}
 \usepackage{amssymb}
 \usepackage{latexsym} 
 \usepackage{amsfonts} 
 \usepackage{epsfig}
 \usepackage{psfrag}

 \newcommand{\nn}{\nonumber\\} 
  
 \newcommand{\f}[1]{\mbox{\boldmath$#1$}}
 
 \newcommand{\vau}{\mbox{\boldmath$v$}}
 \newcommand{\na}{\mbox{\boldmath$\nabla$}}
 \newcommand{\bea}{\begin{eqnarray}}
 \newcommand{\ea}{\end{eqnarray}}
 \newcommand{\eea}{\end{eqnarray}}
 \newcommand{\ord}{{\cal O}}

\begin{document}
  
\title{Quantum backreaction in dilute Bose-Einstein condensates} 

\author{Ralf Sch\"utzhold, Michael Uhlmann, and Yan Xu}

\affiliation{Institut f\"ur Theoretische Physik, 
Technische Universit\"at Dresden, D-01062 Dresden, Germany}

\author{Uwe R.~Fischer}

\affiliation{Eberhard-Karls-Universit\"at T\"ubingen, 
Institut f\"ur Theoretische Physik\\  
Auf der Morgenstelle 14, D-72076 T\"ubingen, Germany}

\begin{abstract} 
For many physical systems which can be approximated by a classical 
background field plus small (linearized) quantum fluctuations, 
a fundamental question concerns the correct description of the 
backreaction of the quantum fluctuations onto the dynamics of the 
classical background.
We investigate this problem for the example of dilute atomic/molecular 
Bose-Einstein condensates, for which the microscopic dynamical behavior 
is under control.
It turns out that the effective-action technique does not yield the 
correct result in general and that the knowledge of the 
pseudo-energy-momentum tensor ${\langle\hat T_{\mu\nu}\rangle}$ 
is not sufficient to describe quantum backreaction.  
\end{abstract}

  
\maketitle

\section{Introduction}\label{Introduction}

Many phenomena in physics can be described, to a sufficient 
degree of accuracy, by means of the background-field method, 
wherein one considers linearized quantum fluctuations on top 
of an approximately classical background. 
For a scalar gap-less (e.g., Goldstone) mode~$\phi$, propagating in an 
arbitrary background (e.g., zeroth sound in inhomogeneous superfluids), 
the low-energy effective action can be cast into the pseudocovariant
form 
\bea
\label{low-energy}
{\cal L}_{\rm eff}=\frac{1}{2}\,(\partial_\mu\phi)
G^{\mu\nu}(\partial_\nu\phi)
\,,
\ea
with ${\partial_\mu=(\partial/\partial t,\na)}$ denoting the
space-time derivatives (we employ the summation convention), 
and the space-time dependent tensor $G^{\mu\nu}$ characterizing the
background \cite{unruh,visser}.
Hence the propagation of~$\phi$ is completely analogous to that of a
scalar field in a curved space-time with the effective metric
$g^{\mu\nu}_{\rm eff}$ depending on the background tensor via 
${G^{\mu\nu}=|g_{\rm eff}|^{1/2}g^{\mu\nu}_{\rm eff}}$, where   
${g_{\rm eff}={\rm det}[g^{\rm eff}_{\mu\nu}]}$
(with possibly an additional dilaton factor).
This striking analogy allows us to apply many tools and methods 
developed for quantum fields in curved space-times \cite{birrell}, 
and to conclude that associated phenomena like Hawking radiation
\cite{unruh}, super-radiance etc., may occur.
Since the equation of motion for the scalar mode, 
${\Box_{\rm eff}\phi=\nabla_\mu^{\rm eff}g^{\mu\nu}_{\rm eff}
\nabla_\nu^{\rm eff}\phi=0}$, 
is equivalent to covariant energy-momentum balance, 
${\nabla_\mu^{\rm eff}T^{\mu\nu}=0}$,
one such tool is the pseudo-energy-momentum tensor, cf.~\cite{stone}
\bea
\label{pseudo-energy-momentum}
T_{\mu\nu}=(\partial_\mu\phi)(\partial_\nu\phi)-\frac12\,
g_{\mu\nu}^{\rm eff}(\partial_\rho\phi)(\partial_\sigma\phi)
g^{\rho\sigma}_{\rm eff}
\,.
\ea
The identification of effects by analogy, described above 
(including the existence of a pseudo-energy-momentum tensor), 
is of a kinematical nature. 
An important question concerns the dynamics, that is, the 
{\em backreaction} of these quantum fluctuations onto 
the classical background. 
Extending the analogy to curved space-times a bit further, one is
tempted to apply the effective-action method  
(see, e.g., \cite{birrell} and \cite{balbinot}), and, 
since the dependence of the effective action~${\cal A}_{\rm eff}$ on
the degrees of freedom of the background~$\eta$ enters via the
effective metric ${g^{\mu\nu}_{\rm eff}=g^{\mu\nu}_{\rm eff}(\eta)}$, 
to determine the backreaction contribution to their equation of
motion by
\bea
\label{pseudo}
\frac{\delta{\cal A}_{\rm eff}}{\delta\eta}
=
\frac{\delta{\cal A}_{\rm eff}}{\delta g^{\mu\nu}_{\rm eff}}\,
\frac{\delta g^{\mu\nu}_{\rm eff}}{\delta\eta}
=
\frac12\,\sqrt{|g_{\rm eff}|}\,
\left\langle\hat T_{\mu\nu}\right\rangle\,
\frac{\partial g^{\mu\nu}_{\rm eff}}{\partial\eta}
\,.
\ea
The precise meaning of the expectation value of the
pseudo-energy-momentum tensor, ${\langle\hat T_{\mu\nu}\rangle}$, is 
difficult to grasp in general, due to the nonuniqueness of the vacuum 
state in a complicated curved space-time background and the
ultraviolet (UV) renormalization procedure.
Adopting a covariant renormalization scheme, the results for 
${\langle\hat T_{\mu\nu}\rangle}$ can be classified in terms of 
geometrical quantities (cf.~the trace anomaly \cite{birrell}).

However, in calculating the quantum backreaction in that way, 
one is implicitly making two essential assumptions:
firstly, that the leading contributions to the backreaction are 
completely determined by the effective action in Eq.\,(\ref{low-energy}), 
and, secondly, that deviations from the low-energy effective action at 
high energies do not affect the (renormalized) expectation value of the
pseudo-energy-momentum tensor, ${\langle\hat T_{\mu\nu}\rangle}$.
Since the effective covariance in Eq.\,(\ref{low-energy}) is only a
low-energy property, the applicability of a covariant renormalization
scheme is not obvious in general. 
In the following, we critically examine 
the question of whether the two assumptions mentioned above are justified, 
e.g., whether ${\langle\hat T_{\mu\nu}\rangle}$ completely determines the
backreaction of the linearized quantum fluctuations. In oder to address
this question, we shall
consider phonon modes in a well understood superfluid, a dilute
Bose-Einstein condensate.

This paper is organized as follows.
In Section~\ref{Bose-Einstein}, we give a brief introduction to
Bose-Einstein condensates, and introduce a 
particle-number-conserving ansatz for the field operator.
In the subsequent Section~\ref{backreaction}, based on this ansatz, the
backreaction to the full current will be calculated, yielding a different
result than that obtained by the effective action method.
The differences of the two expressions will be investigated in more detail
in Sec.~\ref{Comparison}.
Since the choice of the (classical) background is of particular importance,
we will demonstrate its influence in Sec.~\ref{Classical-background}, where
we consider two different choices for the background yielding two different
expressions for the backreaction.
Afterwards, the failure of the effective action technique is discussed in
more detail in Sec.~\ref{Effective-Action}.
The cutoff dependence of the pseudo-energy-momentum tensor
(\ref{pseudo-energy-momentum}) is addressed in Sec.~\ref{cutoff}.
As a simple example, we consider the influence of the backreaction
contribution to a static 1D condensate in Sec.~\ref{Simple-Example}.

\section{Bose-Einstein Condensates}\label{Bose-Einstein}

In the usual $s$-wave scattering approximation, the many-particle
system is described by the field operator in the Heisenberg picture
(we set ${\hbar=m=1}$)
\bea
\label{Heisenberg}
i\frac{\partial}{\partial t}\hat\Psi=
\left(-\frac{1}{2}\,\na^2+V+
g\,\hat\Psi^\dagger\hat\Psi\right)\hat\Psi
\,,
\ea
where $V$ denotes the one-particle trapping potential, 
and the coupling constant $g$ is related to the $s$-wave scattering
length~$a_s$ via ${g=4\pi a_s}$ (in three spatial dimensions).
In the limit of many particles $N\gg1$, in a finite trap at zero
temperature with almost complete condensation, the full field operator
$\hat\Psi$ can be  represented in terms of the 
particle-number-conserving mean-field {ansatz} \cite{Particle,castin} 
\bea
\label{mean-field}
\hat\Psi=\left(\psi_{\rm c}+\hat\chi+\hat\zeta\right)\hat A/\sqrt{\hat N}
\,,
\ea
with the order parameter~${\psi_{\rm c}=\ord(\sqrt{N})}$, the one-particle
excitations~${\hat\chi=\ord(N^0)}$, and the remaining higher-order
corrections~${\hat\zeta=\ord(1/\sqrt{N})}$. 
The above mean-field {ansatz} can be derived in the dilute-gas limit
by formally setting ${g=\ord(1/N)}$ \cite{castin,1/N,derivation,meanfield};
we shall use this formal definition of the dilute-gas limit in what
follows.
Insertion of Eq.\,(\ref{mean-field}) into (\ref{Heisenberg}) yields the
Gross-Pitaevski\v\i\/ equation \cite{GP} for the order
parameter~$\psi_{\rm c}$  
\bea
\label{GP}
i\frac{\partial}{\partial t}\psi_{\rm c}
&=&
\left(
-\frac{1}{2}\,\na^2+V+g|\psi_{\rm c}|^2
+2g\left\langle\hat\chi^\dagger\hat\chi\right\rangle
\right)\psi_{\rm c}
\nn
&&+
g\left\langle\hat\chi^2\right\rangle\psi_{\rm c}^*
\,,
\ea
the Bogoliubov-de~Gennes equations \cite{BdG} for $\hat\chi$ 
\bea
\label{BdG}
i\frac{\partial}{\partial t}\hat\chi=
\left(
-\frac{1}{2}\,\na^2+V+
2g|\psi_{\rm c}|^2\right)\hat\chi
+g\psi^2_{\rm c}\hat\chi^\dagger
\,,
\ea
and for the remaining higher-order 
corrections ${\hat\zeta=\ord(1/\sqrt{N})}$, 
neglecting $\ord(1/N)$ terms:   
\bea
\label{xi}
i\frac{\partial}{\partial t}\hat\zeta
&\approx&
\left(
-\frac{1}{2}\,\na^2+V+2g|\psi_{\rm c}|^2\right)\hat\zeta
+g\psi^2_{\rm c}\,\hat\zeta^\dagger
\nn
&&+
2g(\hat\chi^\dagger\hat\chi-
\langle\hat\chi^\dagger\hat\chi\rangle)\psi_{\rm c}
+g(\hat\chi^2-\langle\hat\chi^2\rangle)\psi_{\rm c}^*
\,.
\ea
Note that the additional terms 
${2g\langle\hat\chi^\dagger\hat\chi\rangle}$ and 
${g\langle\hat\chi^2\rangle}$ in the 
Gross-Pitaevski\v\i\/ equation~(\ref{GP}) 
ensure that the expectation value of 
${\hat\zeta=\ord(1/\sqrt{N})}$ vanishes in leading order,  
${\langle\hat\zeta\rangle=\ord(1/N)}$, see also
\cite{castin,temperature}. 
Without these additional terms, the mean-field
expansion~(\ref{mean-field}) would still be valid with
${\hat\zeta=\ord(1/\sqrt{N})}$, but without 
${\langle\hat\zeta\rangle=\ord(1/N)}$. 

\section{backreaction}\label{backreaction}

The observation that the Gross-Pitaevski\v\i\/ equation~(\ref{GP}) 
yields an equation correct to leading order ${\cal O}(\sqrt{N})$, using
either $|\psi_{\rm c}|^2$ or 
${|\psi_{\rm c}|^2+2\left\langle\hat\chi^\dagger\hat\chi\right\rangle}$ 
in the first line of (\ref{GP}), hints at the fact that quantum
backreaction effects correspond to next-to-leading order terms and
cannot be derived {\em ab initio} in the above manner without
additional assumptions. 
Therefore, we shall employ an alternative method:
In terms of the exact density and current given by 
\bea
\label{exact}
\varrho=\left\langle\hat\Psi^\dagger\hat\Psi\right\rangle
\,,\;
\f{j}
=
\frac{1}{2i}\,
\left\langle\hat\Psi^\dagger\na\hat\Psi
-{\rm H.c.}
\right\rangle
\,,
\ea
the time-evolution is governed by the equation of continuity for 
$\varrho$ and an Euler type equation for the current~$\f{j}$.
After insertion of Eq.\,(\ref{Heisenberg}), we find that the equation 
of continuity is not modified by the quantum fluctuations but
satisfied exactly (i.e., to all orders in $1/N$ or $\hbar$)
\bea
\label{continuity-exact}
\frac{\partial}{\partial t}\varrho+\na\cdot\f{j}=0
\,,
\ea
in accordance with the Noether theorem and the $U(1)$ invariance of 
the Hamiltonian, cf.~\cite{balbinot}.
However, if we insert the mean-field expansion~(\ref{mean-field}) and 
write the full density as a sum of condensed and noncondensed parts 
\bea
\label{rho-chi}
\varrho=
\varrho_{\rm c}+\left\langle\hat\chi^\dagger\hat\chi\right\rangle
+\ord(1/\sqrt{N})
\,,
\ea
with $\varrho_{\rm c}=|\psi_{\rm c}|^2$, we find that neither part is
conserved separately in general. 
Note that this split requires ${\langle\hat\zeta\rangle=\ord(1/N)}$,  
i.e., the modifications to the Gross-Pitaevski\v\i\/ equation~(\ref{GP}) 
discussed above.
Similarly, we may split up the full current 
[with ${\varrho_{\rm c}\vau_{\rm c}=\Im(\psi_{\rm c}^*\na\psi_{\rm c})}$]
\bea
\label{j-chi}
\f{j}=\varrho_{\rm c}\vau_{\rm c}+
\frac{1}{2i}\,
\left\langle\hat\chi^\dagger\na\hat\chi
-{\rm H.c.}
\right\rangle
+\ord(1/\sqrt{N})
\,,
\ea
and introduce an average velocity $\vau$ via ${\f{j}=\varrho\vau}$. 
This enables us to {\em unambiguously} define the quantum
backreaction $\f{Q}$ as the following additional contribution in an
equation of motion for $\f{j}$ analogous to the Euler equation:   
\bea
\label{Euler}
\frac{\partial}{\partial t}\f{j}=
\f{f}(\f{j},\varrho)
+\f{Q}+\ord(1/\sqrt{N})
\,,
\ea
where ($\f{j} = \varrho \vau$)
\bea
\f{f}(\f{j},\varrho) & = & - \vau\left[\na \cdot (\varrho\vau)\right]
        - \varrho( \vau \cdot\na)\vau
        \nonumber\\
        && + \varrho\na\left(\frac{1}{2}
                \frac{\na^2 \sqrt{\varrho}}{\sqrt{\varrho}}
                - V_{\rm ext} - g \varrho \right)
\ea
denotes the usual classical force density term. 
Formulation in terms of the conventional Euler equation yields
\bea
\label{convective}
\left(\frac{\partial}{\partial t}+\f{v}\cdot\na\right)\vau
&=&
-\na\left(
V+g\varrho
-\frac{1}{2}\frac{\na^2\sqrt{\varrho}}{\sqrt{\varrho}}
\right)
\nn
&&
+\f{Q}/\varrho
+\ord(1/\sqrt{N^3})
\,.
\ea
The quantum backreaction $\f{Q}$ can now be calculated
by comparing the two equations above and expressing 
${\partial\f{j}/\partial t}$ in terms of the field operators via 
Eqs.\,(\ref{Heisenberg}) and (\ref{exact})
\bea
\label{field-operators}
\frac{\partial}{\partial t}\f{j}
&=&
\frac14\left\langle\hat\Psi^\dagger\na^3\hat\Psi-
(\na^2\hat\Psi^\dagger)\na\hat\Psi+{\rm H.c.}\right\rangle
\nn
&&
-\left\langle\hat\Psi^\dagger\hat\Psi\right\rangle\na V 
-\frac{1}{2g}\na\left\langle g^2(\hat\Psi^\dagger)^2\hat\Psi^2\right\rangle
\,.
\ea
After insertion of the mean-field expansion~(\ref{mean-field}),  
we obtain the leading contributions in the Thomas-Fermi limit  
\bea
\label{TF}
\f{Q}
&=&
\na\cdot\left(\vau\otimes\f{j}_\chi+\f{j}_\chi\otimes\vau-
\varrho_\chi\vau\otimes\vau\right)
\nn
&&-
\frac{1}{2g}\na\left(g^2
\left\langle
2|\psi_{\rm c}|^2\hat\chi^\dagger\hat\chi+
\psi_{\rm c}^2(\hat\chi^\dagger)^2+(\psi_{\rm c}^*)^2\hat\chi^2
\right\rangle
\right)
\nn
&&
-\frac12\,\na\cdot
\left\langle(\na\hat\chi^\dagger)\otimes\na\hat\chi+{\rm H.c.}\right\rangle
\,,
\ea
with ${\varrho_\chi=\left\langle\hat\chi^\dagger\hat\chi\right\rangle}$
and ${\f{j}_\chi=\Im\left\langle\hat\chi^\dagger\na\hat\chi\right\rangle}$.
Under the assumption that the relevant length scales $\lambda$ 
for variations of, e.g., $\varrho$ and $g$, are much larger than the
healing length ${\xi=(g\varrho_{\rm c})^{-1/2}}$, we have neglected
terms containing quantum pressure contributions $\na^2\varrho$ and
$[\na\varrho]^2$ (Thomas-Fermi or local-density approximation). 
These contributions would, in particular, spoil the effective (local)
geometry in Eq.\,(\ref{low-energy}). 

\section{Comparison with effective-action technique}\label{Comparison}

In order to compare the expression (\ref{TF}) with Eq.\,(\ref{pseudo}),
we have to identify $\phi$ and $G^{\mu\nu}$. 
Phonon modes with wavelength ${\lambda\gg\xi}$ are described by the action
in Eq.\,(\ref{low-energy}) in terms of the phase fluctuations~$\phi$
provided that $G^{\mu\nu}$ is given by 
\bea
\label{PGL}
G^{\mu\nu}
=
\frac{1}{g}
\left( 
\begin{array}{cc} 
1 \, &\, {\vau} \\
{\vau} \, &\, {\vau}\otimes{\vau}-c^2\f{1}
\end{array} 
\right)
\,,
\ea
where $c^2=g\varrho$ is the speed of sound, cf.~\cite{unruh,visser}. 
The density fluctuations~$\delta\varrho$ are related to the phase 
fluctuations~$\phi$ via 
${\delta\varrho=-g^{-1}(\partial/\partial t+\vau\cdot\na)\phi}$.

If we now calculate the quantum corrections to the Bernoulli equation 
using the effective-action method in Eq.~(\ref{pseudo}) by inserting 
$\partial G^{\mu\nu}/\partial\varrho$ or 
$\partial g^{\mu\nu}_{\rm eff}/\partial\varrho$, we obtain
\bea
\label{corrections}
\frac{\delta{\cal A}_{\rm eff}}{\delta\varrho}
=
-\frac12\langle(\na\hat\phi)^2\rangle
\,.
\ea
Clearly, this result differs from the expression~(\ref{TF}) derived in the 
previous Section.
Moreover, it turns out that Eq.\,(\ref{TF}) contains contributions which
are not part of the pseudo-energy-momentum tensor 
${\langle\hat T_{\mu\nu}\rangle}$:
Even though the phonon flux $\f{j}_\chi$ in the first line of 
Eq.\,(\ref{TF}) corresponds to 
${\langle\delta\hat\varrho\na\hat\phi+{\rm H.c.}\rangle}$, 
which indeed reproduces the mixed space-time components of the 
pseudo-energy-momentum tensor ${\langle\hat T_{\mu\nu}\rangle}$,
the phonon density $\varrho_\chi$ contains
${\langle(\delta\hat\varrho)^2\rangle_{\rm ren}}$
(where $\langle\dots\rangle_{\rm ren}$ 
means that the divergent c-number 
${\hat\chi\hat\chi^\dagger-\hat\chi^\dagger\hat\chi=\delta(0)}$
has been subtracted already) which is part of  
${\langle\hat T_{\mu\nu}\rangle_{\rm ren}}$ but also 
${\langle\hat\phi^2\rangle_{\rm ren}}$ 
which is not.
(Note that ${\langle\hat\phi^2\rangle_{\rm ren}}$ cannot be cancelled 
by the other contributions.) 
The expectation value in the second line of Eq.\,(\ref{TF}) is
basically ${\langle(\delta\hat\varrho)^2\rangle_{\rm ren}}$
and thus part of ${\langle\hat T_{\mu\nu}\rangle_{\rm ren}}$.
Finally, the expression 
${\langle(\na\hat\chi^\dagger)\otimes\na\hat\chi+{\rm H.c.}\rangle}$
in the last line of Eq.\,(\ref{TF}) contains 
${\langle\na\hat\phi\otimes\na\hat\phi\rangle_{\rm ren}}$
which does occur in ${\langle\hat T_{\mu\nu}\rangle_{\rm ren}}$, but also 
${\langle\na\delta\hat\varrho\otimes\na\delta\hat\varrho\rangle_{\rm ren}}$,
which does not.
One could argue that the latter term ought to be neglected in the 
Thomas-Fermi or 
local-density approximation since it is on the same footing as the 
quantum pressure contributions $\na^2\varrho$ and  $[\na\varrho]^2$, 
but it turns out that this expectation value yields cutoff dependent 
contributions of the same order of magnitude as the other terms, see 
Section~\ref{cutoff} below.

\section{Choice of classical background}\label{Classical-background}

In order to understand the disagreement between expressions (\ref{TF}) and
(\ref{corrections}), and to demonstrate the dependence of the
explicit expression for the quantum backreaction on the choice of the
classical background (see also Appendix~\ref{Toy Model}), we will consider 
two particular alternative backgrounds.
Firstly, the background is prescribed by the density and current generated
by the condensed part $\psi_{\rm c}$ of the field operator $\hat
\Psi$, 
cf.\ Eqs.~(\ref{rho-chi}) and (\ref{j-chi}).
And secondly, we consider (the expectation values of) the full density and
the velocity potential as a (classical) background.

In the first case, the background current is given by
\bea
\f{j}_{\rm c}
        = \varrho_{\rm c} \vau_{\rm c}
        = \frac{1}{2i}
        \left( \psi_{\rm c}^* \na \psi_{\rm c} - {\rm H.c.} \right)
        = \f{j} - \f{j}_\chi
        \,.
\ea
Note that the exact splitting $\f{j}=\f{j}_{\rm c}+\f{j}_\chi$ is not unique
and thus the classical background $\f{j}_{\rm c}$ and the fluctuation
$\f{j}_\chi$ are (in contrast to $\f{j}$) not directly measurable, cf.~the
remark on the Gross-Pitaevski\v\i\/ equation~(\ref{GP}). 
The evolution of the phonon flux can be inferred from the 
Bogoliubov-de~Gennes equation~(\ref{BdG}) 
\bea
\frac{\partial}{\partial t}
\f{j}_\chi
&=&
\frac14\na^3\varrho_\chi
-
\frac12\,\na\cdot
\left\langle(\na\hat\chi^\dagger)\otimes\na\hat\chi+{\rm H.c.}\right\rangle
\nn
&&
-\varrho_\chi\na\left(V+2\varrho_{\rm c}g\right)
\nn
&&
-\frac12
\left[ 
\left\langle\hat\chi^2\right\rangle 
\na\left(g\left(\psi_{\rm c}^{*}\right)^2\right)
+{\rm H.c.}
\right]   
\,,
\ea
which enables us to derive the evolution equation for the condensate part
via the subtraction $\f{j}_{\rm c}=\f{j}-\f{j}_\chi$
\bea
\label{cond-bern}
\frac{\partial}{\partial t}
\f{j}_{\rm c}
&=&
\f{f}(\varrho_{\rm c},\vau_{\rm c})
-2\varrho_{\rm c}\na(g\varrho_\chi)
\nn
&&
-\frac12
\left[ 
\left(\psi_{\rm c}^{*}\right)^2\na
\left(g\left\langle\hat\chi^2\right\rangle\right)
+{\rm H.c.}
\right]   
\,,
\ea
which is the same result as obtained from the modified
Gross-Pitaevski\v\i\/ equation~(\ref{GP}). 
However, the interpretation of the r.h.s.\ of the equation above as a force 
density is not straightforward, since the equation of continuity is not 
satisfied
\bea
\frac{\partial}{\partial t}\varrho_{\rm c}+\na\cdot\f{j}_{\rm c}=
ig\left\langle
\psi_{\rm c}^2\left(\hat{\chi}^{\dagger}\right)^2-
\left(\psi_{\rm c}^{*}\right)^2\hat\chi^2
\right\rangle   
\,.
\ea
I.e., an acceleration of the atoms of the condensate caused by a real 
force could make them leave the condensate or induce a change in 
$\f{j}_{\rm c}$ and it is difficult to disentangle these two effects.
Roughly speaking, choosing the condensate part $\f{j}_{\rm c}$ 
(which is not measurable and not conserved) as the classical background 
would correspond to the artificial variable $Y$ of the toy model discussed
in Appendix~\ref{Toy Model}, whereas the full (measurable and conserved)
current $\f{j}$ corresponds to $X$.

In the second case, we use exactly the same variables as employed in the 
effective-action method described earlier and take the expectation values 
of the operators corresponding to the density and the velocity potential 
as background
\bea
\label{split}
\hat\Phi
&=&
\langle\hat\Phi\rangle+\hat\phi=\phi_{\rm b}+\hat\phi
\,,
\nn
\hat\varrho
&=&
\langle\hat\varrho\rangle+\delta\hat\varrho=\varrho_{\rm b}+\delta\hat\varrho
\,.
\ea
Note that, in contrast to the full density which is a well-defined and 
measurable quantity, the velocity potential, $\hat\Phi$, is not
\cite{Froehlich}.
It can be introduced via the following ansatz for the full field operator
\bea
\label{Madelung}
\hat\Psi=e^{i\hat\Phi}\,\sqrt{\hat\varrho}
\,.
\ea
Since $\hat\Phi$ and $\hat\varrho$ do not commute, other forms such as 
${\hat\Psi=\sqrt{\hat\varrho}\,e^{i\hat\Phi}}$ would not generate a 
self-adjoint $\hat\Phi$ (and simultaneously satisfy 
$\hat\Psi^\dagger\hat\Psi=\hat\varrho$).
Insertion of the above ansatz into the expression for the full current 
yields
\bea
\f{j}=\left\langle\sqrt{\hat\varrho}\;\na\hat\Phi\,\sqrt{\hat\varrho}\,
\right\rangle
\,.
\ea
The quantum corrections to the equation of continuity in this formulation 
can be obtained by inserting the split in Eq.~(\ref{split}) and neglecting
terms of third or higher order
\bea
\f{j}=\varrho_{\rm b}\na\phi_{\rm b}+
\frac12
\langle\delta\hat\varrho\na\hat\phi+{\rm H.c.}\rangle
=\varrho_{\rm b}\na\phi_{\rm b}+\f{j}_\chi
\,,
\ea
and are in perfect agreement with the effective-action method
$\delta{\cal A}_{\rm eff}/\delta\phi_{\rm b}$.

In order to relate the background velocity $\na\phi_{\rm b}=\vau_{\rm b}$
to known quantities (the background density of course equals the full 
density $\varrho_{\rm b}=\varrho$), we remember 
$\f{j}=\f{j}_{\rm c}+\f{j}_\chi$ and deduce
\bea
\varrho_{\rm b}\na\phi_{\rm b}
=
\varrho_{\rm b}\vau_{\rm b}
=
\varrho_{\rm c}\vau_{\rm c}
\,,
\ea
i.e., the background velocity is 
(again up to terms of third or higher order)
proportional to the condensate velocity.
Now, considering a condensate which is initially at rest 
$\vau_{\rm c}(t=0)=0$ and calculating 
\bea
\frac{\partial}{\partial t}\vau_{\rm b}
&=&
\frac{\varrho_{\rm c}}{\varrho_{\rm b}}\,
\frac{\partial}{\partial t}\vau_{\rm c}
=
\frac{\partial _t\f{j}_{\rm c}}{\varrho_{\rm c}+\varrho_\chi}
\nn
&=&
\frac{\f{f}(\varrho_{\rm c}=\varrho_{\rm b}-\varrho_\chi,\vau_{\rm c}=0)}
{\varrho_{\rm c}+\varrho_\chi}
-2\na(g\varrho_\chi)
\nn
&&
-\frac1{2\varrho_{\rm c}}
\left[ 
\left(\psi_{\rm c}^{*}\right)^2\na
\left(g\left\langle\hat\chi^2\right\rangle\right)
+{\rm H.c.}
\right]   
\,,
\ea
we observe that $\partial_t\vau_{\rm b}$ contains the terms $\varrho_\chi$, 
$\langle\hat\chi^2\rangle$, and $\langle(\hat\chi^\dagger)^2\rangle$, 
but not $\langle(\na\hat\phi)^2\rangle$.
As a result, the quantum corrections to the Bernoulli equation
obtained in the above direct manner are in contradiction to the
effective-action method result 
$\delta{\cal A}_{\rm eff}/\delta\varrho_{\rm b}$.

\section{Failure of Effective-Action Technique}\label{Effective-Action}

After having demonstrated the failure of the effective-action method for 
deducing the quantum backreaction, let us study the reasons for this 
failure in more detail.
The full action governing the dynamics of the fundamental fields
$\Psi^*$ and $\Psi$ reads
\bea
{\cal L}^\Psi
=
i\Psi^*\frac{\partial}{\partial t}\Psi
-
\Psi^*\left(-\frac{1}{2}\,\na^2+V+\frac{g}{2}\,\Psi^*\Psi\right)\Psi
\,.
\ea
Linearization $\Psi=\psi_{\rm c}+\chi$ yields the effective second-order 
action generating the Bogoliubov-de~Gennes equations~(\ref{BdG})
\bea
{\cal L}_{\rm eff}^\chi
&=&
i\chi^*\frac{\partial}{\partial t}\chi
-
\chi^*\left(-\frac{1}{2}\,\na^2+V+2g\,|\psi_{\rm c}|^2\right)\chi
\nn
&&-\left[\frac{g}{2}\,(\psi^*_{\rm c})^2\chi^2+{\rm H.c.}\right]
\,.
\ea
Now $\delta{\cal A}_{\rm eff}^\chi/\delta\varrho_{\rm c}$ indeed yields 
the correct backreaction~(\ref{cond-bern}) in the condensate formulation.

However, if we start with the action in terms of the nonfundamental 
variable $\Phi$
\bea
{\cal L}
=
-\varrho\left(\frac{\partial}{\partial t}\Phi+\frac12(\na\Phi)^2\right)
-\epsilon[\varrho] -V\varrho 
\,,
\ea
with $\epsilon[\varrho]$ denoting the internal energy density,
the quantum corrections to the equation of continuity 
$\delta{\cal A}_{\rm eff}/\delta\phi_{\rm b}$ are reproduced correctly 
but the derived quantum backreaction contribution to the Bernoulli 
equation $\delta{\cal A}_{\rm eff}/\delta\varrho_{\rm b}$ is wrong.

Why is the effective-action method working for the fundamental field
$\Psi$, but not for the nonfundamental variable $\Phi$?
The quantized fundamental field $\hat\Psi$ satisfies the equation of 
motion (\ref{Heisenberg}) as derived from the above action and possesses
a well-defined linearization via the mean-field expansion (\ref{mean-field}).
One of the main assumptions of the effective-action method is a similar 
procedure for the variable $\Phi$, i.e., the existence of a well-defined 
and linearizable full quantum operator $\hat\Phi$ satisfying the Bernoulli 
equation (for large length scales)
\bea
\frac{\partial}{\partial t}\hat\Phi+\frac12(\na\hat\Phi)^2
+h[\hat\varrho]
\stackrel{?}{=}0
\,,
\ea
where $h=d\epsilon/d\varrho+V$.
The problem is that the commutator of $\hat\varrho$ and $\hat\Phi$ at the 
same position diverges and hence the quantum Madelung ansatz in 
Eq.~(\ref{Madelung}) is singular.
As a result, the above quantum Bernoulli equation is not well-defined 
(in contrast to the equation of continuity), i.e., insertion of the 
quantum Madelung ansatz in Eq.~(\ref{Madelung}) into Eq.~(\ref{Heisenberg})
generates divergences \cite{Froehlich}.

In order to study these divergences by means of a simple example, 
let us consider the generalized Bose-Hubbard Hamiltonian
\bea
\hat H
=
\alpha\sum\limits_{<ij>}(\hat\Psi^\dagger_i\hat\Psi_j+{\rm H.c.})
+
\sum\limits_i(\beta\hat n_i+\gamma\hat n_i^2)
\,,
\ea
where $<ij>$ denote nearest neighbors and 
$\hat n_i=\hat\Psi^\dagger_i\hat\Psi_i$ the filling factor of one 
lattice site $i$.
In the usual continuum limit, this Hamiltonian generates 
Eq.~(\ref{Heisenberg}) where $\alpha$ is related to the lattice
spacing and the effective mass, $\gamma$ determines $g$, and $V$ is
governed by $\alpha$, $\beta$, and $\gamma$.
Inserting the quantum Madelung ansatz in Eq.~(\ref{Madelung}), however,
the problem of operator ordering arises and the (for the Bernoulli 
equation) relevant term reads
\bea
\hat H_\Phi
=
\frac14\sum\limits_i\sqrt{\hat n_i(\hat n_i+1)}\;(\na\hat\Phi)^2_i+{\rm H.c.}
\,,
\ea
with the replacement $\hat n_i+1$ instead of $\hat n_i$ being one 
effect of the noncommutativity.
In the superfluid phase with large filling $n\gg1$, we obtain the following 
leading correction to the equation of motion 
\bea
\frac{\partial}{\partial t}\hat\Phi+\frac12(\na\hat\Phi)^2
+h[\hat\varrho]
+\frac{1}{\hat n}\,\frac{(\na\hat\Phi)^2}{16}\,\frac{1}{\hat n}
=
\ord\left(\frac{1}{n^3}\right)
\,,
\ea
which depends on microscopic details (such as the filling).

By means of this simple example, we already see that the various limiting 
procedures such as 
the quantization and subsequent mean-field expansion,
the variable transformation $(\Psi^*,\Psi)\leftrightarrow(\varrho,\Phi)$,
and the linearization for small fluctuations, 
as well as continuum limit
do not commute in general -- which explains the failure of the 
effective-action method for deducing the quantum backreaction.
The variable transformation $(\Psi^*,\Psi)\leftrightarrow(\varrho,\Phi)$
is applicable to the zeroth-order equations of motion for the classical 
background as well as to the first-order dynamics of the linearized 
fluctuations -- but the quantum backreaction is a second-order effect, 
where the aforementioned difficulties arise \cite{Froehlich}.

\section{cutoff Dependence}\label{cutoff}

As mentioned in the Introduction, another critical issue is the UV 
divergence of ${\langle\hat T_{\mu\nu}\rangle}$.
Extrapolating the low-energy effective action in
Eq.\,(\ref{low-energy}) to large momenta $k$, the expectation values 
${\langle\delta\hat\varrho^2\rangle}$ and 
${\langle\hat\phi^2\rangle}$ entering $\varrho_\chi$ would diverge. 
For Bose-Einstein condensates, we may infer the deviations from
Eq.\,(\ref{low-energy}) at large $k$ from the Bogoliubov-de~Gennes
equations~(\ref{BdG}). 
Assuming a static and homogeneous background (which should be a good
approximation for large~$k$), a normal-mode expansion yields
\bea
\label{normal-mode}
\hat\chi_{\boldsymbol k}
=
\sqrt{\frac{{\boldsymbol k}^2}{2\omega_{\boldsymbol k}}}
\left[
\left(\frac{\omega_{\boldsymbol k}}{{\boldsymbol k}^2}-\frac12\right)
\hat a_{\boldsymbol k}^\dagger
+
\left(\frac{\omega_{\boldsymbol k}}{{\boldsymbol k}^2}+\frac12\right)
\hat a_{\boldsymbol k}
\right]
\,,
\ea
where $\hat a_{\boldsymbol k}^\dagger$ and $\hat a_{\boldsymbol k}$
are the creation and annihilation operators of the phonons,
respectively, and the frequency~$\omega_{\boldsymbol k}$ is determined 
by the Bogoliubov dispersion relation 
${\omega_{\boldsymbol k}^2=g\varrho\,{\boldsymbol k}^2+{\boldsymbol k}^4/4}$.
Using a linear dispersion 
${\omega_{\boldsymbol k}^2\propto{\boldsymbol k}^2}$ instead, 
expectation values such as 
${\langle\hat\chi^\dagger\hat\chi\rangle}$ 
would be UV divergent, but the correct dispersion relation implies  
${\hat\chi_{\boldsymbol k}\sim
\hat a_{\boldsymbol k}^\dagger\,g\varrho/{\boldsymbol k}^2+
\hat a_{\boldsymbol k}}$ for large ${\boldsymbol k}^2$, and hence 
${\langle\hat\chi^\dagger\hat\chi\rangle}$ is UV finite in three and
lower spatial dimensions.
Thus the healing length $\xi$ acts as an effective UV cutoff 
$k^{\rm cut}_\xi$. 

Unfortunately, the quadratic decrease for large~$k$ in
Eq.\,(\ref{normal-mode}), 
${\hat\chi_{\boldsymbol k}\sim
\hat a_{\boldsymbol k}^\dagger\,g\varrho/{\boldsymbol k}^2+
\hat a_{\boldsymbol k}}$, 
is not sufficient for rendering the other expectation values 
(i.e., apart from $\varrho_\chi$ and $\f{j}_\chi$)
in Eq.\,(\ref{TF}) UV finite in three spatial dimensions.
This UV divergence indicates a failure of the $s$-wave
pseudo-potential ${g\delta^3(\f{r}-\f{r'})}$ in Eq.\,(\ref{Heisenberg}) 
at large wave-numbers~$k$ and can be eliminated by replacing
${g\delta^3(\f{r}-\f{r'})}$ by a more appropriate two-particle
interaction potential ${V_{\rm int}(\f{r}-\f{r'})}$, see \cite{Lee}. 
Introducing another UV cutoff wavenumber~$k^{\rm cut}_s$ related to
the breakdown of the $s$-wave pseudopotential, we obtain 
${\langle(\na\hat\chi^\dagger)\otimes\na\hat\chi+{\rm H.c.}\rangle
\sim g^2\varrho^2\,k^{\rm cut}_s}$
and ${\langle\hat\chi^2\rangle\sim g\varrho\,k^{\rm cut}_s}$.

In summary, there are two different cutoff wavenumbers:
The first one, $k^{\rm cut}_\xi$, is associated to the breakdown of the 
effective Lorentz invariance (change of dispersion relation from linear 
to quadratic) and renders some -- but not all -- of the naively 
divergent expectation values finite.
The second wavenumber, $k^{\rm cut}_s$, describes the cutoff for all 
(remaining) UV divergences.
In dilute Bose-Einstein condensates, these two scales are vastly 
different by definition; because the system is dilute, the inverse
range of the true potential, which is of order $k^{\rm cut}_s$, must
be much larger than the inverse healing length:  
\bea
k^{\rm cut}_{\rm UV}
=
k^{\rm cut}_s
\gg
k^{\rm cut}_\xi
=
k^{\rm cut}_{\rm Lorentz}
\,.
\ea
Note the opposite relation 
$k^{\rm cut}_{\rm Lorentz} \gg k^{\rm cut}_{\rm UV}$
is very unnatural since every quantum field theory which 
has the usual properties such as locality and Lorentz invariance etc.,
must have UV divergences (e.g., in the two-point function).

The renormalization of the cutoff-dependent terms is different for the 
two cases:
The $k^{\rm cut}_s$-contributions can be absorbed by a 
$\varrho$-independent renormalization of the coupling $g$ 
\cite{meanfield,Lee}, whereas the 
$k^{\rm cut}_\xi$-contributions depend on the density in a nontrivial 
way and thus lead to a renormalization of the pressure and the chemical 
potential etc., see the next Section.

\section{Simple Example}\label{Simple-Example}

In order to provide an explicit example for the quantum backreaction
term in Eq.\,(\ref{TF}), without facing the UV problem, let us consider 
a quasi-one-dimensional (quasi-1D) condensate 
\cite{1D,Goerlitz}, where all the involved quantities are UV finite. 
In accordance with general considerations \cite{Uncertainty},
the phonon density $\varrho_\chi$ is infrared (IR) divergent
in one spatial dimension, therefore inducing finite-size effects.  
Nevertheless, in certain situations, we are able to derive a closed 
local expression for the quantum backreaction term $Q$:
Let us assume a completely static condensate ${\vau=0}$ 
in effectively one spatial dimension, still allowing for a spatially
varying density $\varrho$ and possibly also coupling $g$.
Furthermore, since spatial variations of $\varrho$ and $g$ occur 
on length scales $\lambda$ much larger than the healing length, 
we keep only the leading terms in $\xi/\lambda\ll 1$, i.e., the
variation of $\varrho$ and $g$ will be neglected in the calculation of
the expectation values. 
In this case, the quantum backreaction term $Q$ 
simplifies considerably and yields 
(in effectively one spatial dimension, where ${g\equiv g_{\rm 1D}}$
and ${\varrho\equiv\varrho_{\rm 1D}}$ now both refer to the 1D
quantities)  
\bea
\label{1+1}
Q
&=&
-
\nabla\left\langle(\nabla\hat\chi^\dagger)\nabla\hat\chi\right\rangle
-
\frac{1}{2g}\nabla\left(g^2\varrho\left\langle
2\hat\chi^\dagger\hat\chi+(\hat\chi^\dagger)^2+\hat\chi^2
\right\rangle
\right)
\nn
&=&
-\nabla\left(\frac{1}{3\pi}\,(g\varrho)^{3/2}\right)
+\frac{1}{2\pi g}\nabla\left(g^{5/2}\varrho^{3/2}\right)
+\ord(\xi^2/\lambda^2)
\nn
&=&
\frac{\varrho}{2\pi}\,\nabla\sqrt{g^3\varrho}
+\ord(\xi^2/\lambda^2)
\,.
\ea
It turns out that the IR divergences of 
${2\langle\hat\chi^\dagger\hat\chi\rangle}$ and 
${\langle(\hat\chi^\dagger)^2+\hat\chi^2\rangle}$ cancel each other 
such that the resulting expression is not only UV but also IR finite.  
Note that the sign of~$Q$ is positive and hence opposite to the contribution 
of the pure phonon density ${\langle\hat\chi^\dagger\hat\chi\rangle}$, 
which again illustrates the importance of the term 
${\langle(\hat\chi^\dagger)^2+\hat\chi^2\rangle}$.

A possible experimental signature of the quantum
backreaction term $Q$ calculated above, is the change
incurred on the static Thomas-Fermi solution of the
Euler equation~(\ref{convective}) for the density distribution 
(cf.~\cite{GP,Lee})
\bea
\label{exp}
\varrho_{\rm 1D}
=
\frac{\mu-V}{g_{\rm 1D}}+\frac{\sqrt{\mu-V}}{2\pi}
+\ord(1/\sqrt{N})
\,,
\ea
with $\mu$ denoting the (constant) chemical potential. 
The classical [$\ord(N)$] density profile 
${\varrho_{\rm cl}=(\mu-V)/g_{\rm 1D}}$ acquires nontrivial quantum 
[$\ord(N^0)$] corrections ${\varrho_Q=\sqrt{\mu-V}/2\pi}$ 
where the small parameter is the ratio of the interparticle distance 
${1/\varrho=\ord(1/N)}$ over the healing 
length~${\xi=\ord(N^0)}$.
Note that the quantum backreaction term
${\varrho_Q}$ in the above split 
${\varrho=\varrho_{\rm cl}+\varrho_Q}$
should neither be confused with the phonon density $\varrho_\chi$ in  
${\varrho=\varrho_{\rm c}+\varrho_\chi}$ (remember that $\varrho_\chi$ is IR 
divergent and hence contains finite-size effects) nor
with the quantum pressure contribution $\propto\na^2\sqrt{\varrho}$ in  
Eq.\,(\ref{convective}). 

Evaluating explicitly the change $\Delta R$ of the Thomas-Fermi size  
(half the full length), where ${\mu=V}$, of a quasi-1D  
Bose-Einstein condensate induced by backreaction, 
from Eq.\,(\ref{exp}) we get 
${\Delta R=-2^{-5/2}(\omega_\perp/\omega_z)a_s}$. 
(The quasi-1D coupling constant $g_{\rm 1D}$ is related to the 3D
$s$-wave scattering length $a_s$ and the perpendicular harmonic
trapping $\omega_\perp$ by ${g_{\rm 1D}=2a_s\omega_\perp}$
\cite{1D}.) 
In units of the classical size 
$R_{\rm cl}=(3a_sN\omega_\perp/\omega_z^2)^{1/3}$, 
we have     
\begin{equation}
\label{radius}
\frac{\Delta R}{R_{\rm cl}}
= 
-\frac{1}{4\sqrt2} 
\left(\frac{1}{3N}\right)^{1/3} 
\left(\frac{\omega_\perp}{\omega_z}
\frac{a_s}{a_z}\right)^{2/3}, 
\end{equation}   
where $a_z=1/\sqrt{\omega_z}$ describes the longitudinal harmonic
trapping. 
In quasi-1D condensates, backreaction thus leads to a
{\em shrinking} of the cloud relative to the classical expectation -- 
whereas in three spatial dimensions we have the opposite effect
\cite{GP,Lee}. 
For reasonably realistic experimental parameters, the effect of
backreaction should be measurable; for 
${N\simeq 10^3}$, ${\omega_\perp/\omega_z\simeq 10^3}$, 
and ${a_s/a_z\simeq 10^{-3}}$, 
we obtain ${|\Delta R/{R_{\rm cl}}|\simeq1\%}$.
  
\section{Conclusions}\label{Conclusions}

Even though the explicit form of the quantum backreaction terms depends
on the definition of the classical background, the effective-action method 
does not yield the correct result in the general case (i.e., independent of 
the choice of variable etc.).
The knowledge of the classical (macroscopic) equation of motion -- such as 
the Bernoulli equation -- may be sufficient for deriving the first-order 
dynamics of the linearized quantum fluctuations (phonons), but the quantum 
backreaction as a second-order effect cannot be obtained without further
knowledge of the microscopic structure (e.g., operator ordering).
It is tempting to compare these findings to gravity, where we also know the 
classical equations of motion only 
\bea
R_{\mu\nu}-\frac12\,g_{\mu\nu}\,R=\kappa\,T_{\mu\nu}
\,,
\ea
which -- in analogy to the Bernoulli equation -- might yield the correct
first-order equations of motion for the linearized gravitons, but perhaps
not their (second-order) quantum backreaction.
Another potentially interesting point of comparison is the existence of two 
different high-energy scales -- one associated to the breakdown of Lorentz 
invariance $k^{\rm cut}_\xi=k^{\rm cut}_{\rm Lorentz}$ and the other 
$k^{\rm cut}_{\rm UV}=k^{\rm cut}_s$ to the UV cutoff as well as the 
question of whether one of the two (or both) correspond to the Planck 
scale in gravity.

The dominant ${\ord(\xi/\lambda)}$ quantum backreaction contributions 
like those in Eq.\,(\ref{exp}) depend on the healing length as the lower
UV cutoff and hence cannot be derived from the low-energy effective action 
in Eq.\,(\ref{low-energy}) using a covariant (i.e., cutoff independent)  
regularization scheme, which does not take into account details of
microscopic physics (represented, for example, in the 
quasiparticle dispersion relation). 
Note that the leading ${\ord(\xi/\lambda)}$ quantum correction to the
pressure could be identified with a cosmological term,  
${\langle\hat T_{\mu\nu}\rangle=\Lambda\,g_{\mu\nu}}$ 
in Eq.\,(\ref{pseudo}), provided that $\Lambda$ is not constant but
depends on $g$ and $\varrho$.
(In general relativity, the Einstein equations demand $\Lambda$ to be
constant.) 

As became evident from the remarks after Eq.\,(\ref{PGL}), 
the knowledge of the expectation value of the pseudo-energy-momentum 
tensor ${\langle\hat T_{\mu\nu}\rangle}$ is not sufficient for 
determining the quantum backreaction effects in general.
Even though ${\langle\hat T_{\mu\nu}\rangle}$ is a useful concept
for describing the phonon kinematics (at low energies), and one may 
identify certain contributions in Eq.\,(\ref{TF}) with terms occurring
in ${\langle\hat T_{\mu\nu}\rangle}$ in curved space-times
(e.g., next-to-leading-order terms in $\xi/\lambda$ in the gradient
expansion), we have seen that it does not represent the full dynamics. 
Related limitations of the classical pseudo-energy-momentum tensor
have been discussed in \cite{stone}.  

In general, the quantum backreaction corrections to the Euler
equation in Eq.\,(\ref{convective}) cannot be represented as the 
gradient of some local potential, cf.~Eq.\,(\ref{TF}). 
Hence they may effectively generate vorticity and might serve as the
seeds for vortex nucleation. 
Note that this effect cannot be observed in the ``force'' density 
in the condensate formulation~(\ref{cond-bern}) since 
$\f{j}_{\rm c}=\varrho_{\rm c}\na\phi_{\rm c}$ holds and thus 
introducing vorticity into the condensate requires creating 
condensate holes with $\varrho_{\rm c}=0$ (vortices),
 which is beyond the regime of
applicability of our linearized analysis.
Therefore, vorticity (in the full-$\f{j}$ formulation) can only be
generated by extracting atoms from the condensate.

In contrast to the three-dimensional case (see, e.g., \cite{Lee,GP}), 
the quantum backreaction corrections given by Eq.\,(\ref{1+1}) 
{\em diminish} the pressure in condensates that can be described 
by Eq.~(\ref{Heisenberg}) in one spatial dimension (quasi-1D case). 
This is a direct consequence of the so-called ``anomalous'' term 
${\langle(\hat\chi^\dagger)^2+\hat\chi^2\rangle}$ in Eq.~(\ref{1+1}), 
which -- together with the cancellation of the IR divergence -- 
clearly demonstrates that it cannot be neglected in general. 
We emphasize that even though Eqs.\,(\ref{1+1})-(\ref{radius}) describe
the {\em static} quantum backreaction corrections to the ground
state, which can be calculated by an alternative method~\cite{Lee} as
well, the expression in  Eq.\,(\ref{TF}) is valid for more general
dynamical situations, such as expanding condensates. 
The static quantum backreaction corrections to the ground state can
be absorbed by a redefinition of the chemical potential
$\mu(\varrho)$ determining a (barotropic) equation of state $p(\varrho)$;
this is however not possible for the other terms in Eq.\,(\ref{TF}), 
such as the quantum friction-type terms $\f{j}\otimes\vau$ etc., 
occurring in more general dynamical situations.

We have derived, from the microscopic physics of dilute Bose-Einstein
condensates, the backreaction of quantum fluctuations onto the motion 
of the full fluid and found possible experimentally observable
consequences. 
We observed a failure of the effective-action technique,
Eq.\,(\ref{pseudo}), and a cutoff dependence of the backreaction 
term due to the breakdown of covariance at high energies. 
Whether similar problems beset ``real'' (quantum) gravity remains an
interesting open question. 

\bigskip

\acknowledgments

R.\,S., M.\,U., and Y.\,X. gratefully acknowledge financial support by
the Emmy Noether Programme of the German Research Foundation (DFG)
under grant No.~SCHU~1557/1-1/2.
The authors~acknowledge support by the COSLAB programme of the  
ESF. 

\bigskip

\appendix
\section{Toy Model}\label{Toy Model}

To give an illustration of the influence of the choice of coordinates
describing the classical background, let us consider a very simple quantum 
system, the one-dimensional harmonic oscillator with the frequency $\Omega$.
The equation of motion for the position operator $\hat X$ reads 
\bea
\frac{\partial^2}{\partial t^2}\hat X+\Omega^2\hat X=0
\,.
\ea
Of course (as is well known from the theory of coherent states), 
splitting up this position operator $\hat X$ into a classical 
(background) part $X_{\rm b}=\langle\hat X\rangle$ and a quantum 
fluctuation part $\delta\hat X$ with $\langle\delta\hat X\rangle=0$ 
via $\hat X=X_{\rm b}+\delta\hat X$ yields the same equation of motion
for both parts separately, i.e., there is no backreaction at all.

However, if we artificially introduce another variable $Y$ via $X=Y^2/2$,
its (classical) equation of motion reads 
\bea
\frac{\partial ^2 }{\partial t ^2} Y +  
\frac 1Y \left(\frac{\partial}{\partial t} Y\right)^2 +\frac12 \Omega^2Y=0
\,.
\ea
Now the same procedure $\hat Y=Y_{\rm b}+\delta\hat Y$ would yield 
a nonvanishing quantum backreaction term. 
Therefore, one must be careful when comparing different expressions 
for the quantum backreaction, since the explicit form may depend on
the choice of splitting the system into a classical background and 
quantum fluctuations (e.g., $\hat X=X_{\rm b}+\delta\hat X$ versus 
$\hat Y=Y_{\rm b}+\delta\hat Y$).
This illustrates the importance of working with measurable quantities
such as $\hat X$ or $\f{j}$ and $\varrho$.
Exploiting the simple example a bit further, an an-harmonic oscillator 
$\partial_t^2\hat X+\Omega^2\hat X=g\hat X^2$ yields 
\bea
\frac{\partial^2}{\partial t^2}\langle\hat X\rangle
+ \Omega^2 \langle\hat X\rangle - g \langle\hat X\rangle^2
= g \langle\delta\hat X^2\rangle \,,
\ea
i.e., a real quantum backreaction term.



\begin{thebibliography}{499}
\bibitem{derivation} 
The total particle number operator ${\hat N=\hat A^\dagger\hat A}$ and
the corresponding creation and annihilation operators satisfy 
${[\hat A,\hat A^\dagger]=1}$ and 
${[\hat\chi,\hat A{\hat N}^{-1/2}]=[\hat\zeta,\hat A{\hat N}^{-1/2}]=0}$, 
i.e., the excitations $\hat\chi$ and $\hat\zeta$ are
particle-number-conserving, cf.~\cite{Particle,castin}.
%
The mean-field ansatz in Eq.\,(\ref{mean-field}) can be motivated by
starting with $N$ free particles, $g=0$, in the same single-particle
state $\psi_{\rm c}$ with $\hat\zeta=0$ and subsequently switching on the
coupling $g>0$ by following the evolution in
Eqs.\,(\ref{GP})-(\ref{xi}) such that the corrections
$\hat\zeta=\ord(1/\sqrt{N})$ remain small \cite{meanfield}.
\end{thebibliography}
\end{document}